# Quantum control of a spin qubit coupled to a photonic crystal cavity

Samuel G. Carter[1†], Timothy M. Sweeney[2†], Mijin Kim[3], Chul Soo Kim[1], Dmitry Solenov[2], Sophia E. Economou[1], Thomas L. Reinecke[1], Lily Yang[2], Allan S. Bracker[1], and Daniel Gammon[1]

[1] *Naval Research Laboratory, Washington, DC 20375, USA*
[2] *NRC postdoctoral associate residing at the Naval Research Laboratory, Washington, DC, 20375, USA*
[3] *Sotera Defense Solutions, Inc., Annapolis Junction, MD 20701, USA*
[†] *These authors contributed equally to this work.*

**A key ingredient for a quantum network is an interface between stationary quantum bits and photons, which act as flying qubits for interactions and communication. Photonic crystal architectures are promising platforms for enhancing the coupling of light to solid state qubits. Quantum dots can be integrated into a photonic crystal, with optical transitions coupling to photons and spin states forming a long-lived quantum memory. Many researchers have now succeeded in coupling these emitters to photonic crystal cavities, but there have been no demonstrations of a functional spin qubit and quantum gates in this environment. Here we have developed a coupled cavity-quantum dot system in which the dot is controllably charged with a single electron. We perform the initialization, rotation and measurement of a single electron spin qubit using laser pulses and find that the cavity can significantly improve these processes.**

A solid state optical cavity in a semiconductor photonic crystal (PhC) membrane has many advantages for nanophotonics and cavity quantum electrodynamics (CQED)[1, 2]. It can have a small volume and high quality factor (Q)[3]; it can be combined with waveguides into extended and complex photonic architectures[3-5]; and it can be integrated with semiconductor electronic devices[6-8]. In the last decade there has been rapid progress both in the development of PhCs themselves and in the study of solid state emitters coupled to PhC cavities.[2-13] Both quantum dots (QDs) and nitrogen-vacancy centers can be incorporated into PhC cavities and can have long-lived spin states[14, 15]. This work has led to the vision of a quantum network[16], similar to that being developed for atomic systems[17], but in a scalable solid-state platform[18, 19].

Proposals for quantum networks almost always involve quantum memories with three energy levels in a Λ configuration – *i.e.* with two ground states and an optically excited state[16, 18, 19]. The two ground state spin levels act as a long-lived quantum memory, while the optically excited state serves to connect the ground state spin coherence to optical coherence. Yet almost all experimental studies to date involving QDs in PhC cavities and waveguides have used uncharged QDs that essentially act as short-lived two-level systems. One way to obtain a Λ-type three-level system envisioned for a large scale quantum network is to charge a QD with a single electron. Recently, diodes have been incorporated into PhC membranes[6-8] that can serve this purpose, and one group has demonstrated controlled charging of a QD in a cavity[20].

While optical initialization, readout, and quantum gates have been performed on spin qubits previously, performing these in the presence of a PhC cavity presents advantages and additional challenges. In particular, the photon density of states in a PhC cavity is strongly modified from free space and is polarization dependent. Here we demonstrate complete quantum control of a QD spin qubit coupled to a PhC cavity. We show: (1) controlled charging of a QD in a cavity with a single electron using a diode incorporated in the PhC; (2) the necessary Λ-type three-level system that results from a magnetic field applied transverse to the sample; (3) spin measurement and initialization using resonant laser spectroscopy; and (4) single qubit gates with picosecond optical pulses. The polarization dependent coupling of the cavity to the QD actually has significant advantages for spin measurement and readout, and the single qubit gates are performed by detuning the pulses from the cavity resonance, thus avoiding polarization dependent coupling.

## Charged quantum dot coupled to a cavity

The PhC membrane with defect optical cavity (L3) is patterned and etched into an n-i-p (n-type, intrinsic, p-type) GaAs diode in which InAs QDs are grown in the intrinsic region closer to the n-type layer, as shown schematically in Figure 1a. Charging of the QDs with a controlled number of electrons is performed via a forward bias across the diode. The charge state of a QD is determined from its photoluminescence (PL) as a function of bias. The PL bias map of a QD in a cavity (labeled QD-C1) is displayed in Fig. 1c, showing the neutral exciton ($X^0$), negatively charged exciton ($X^-$), and cavity lines. The $X^-$ line, indicating a single electron in the QD, is stable over a bias from 1.7-1.95V and is detuned from the cavity by 0.2-0.3 meV over this range. The cavity linewidth of ~0.3 meV corresponds to a quality factor (Q) of 4,000.

Resonant laser excitation of the QD and the cavity mode is used to optically address and measure the spin states of the electron. The differential reflectivity[21, 22] ($\Delta R$) of a laser scanned across the $X^-$, cavity, and $X^0$ resonances is displayed in Fig. 1d,e, and is taken with two laser polarizations, parallel (V) and perpendicular (H) to the cavity polarization. The dominant feature at 1301.93meV for V polarization corresponds to the $X^-$ line, which has a dispersive lineshape due to proximity to the cavity mode and interference effects in the membrane. The cavity mode appears at ~1302.1meV as a broad (~0.3 meV) feature also with a highly dispersive lineshape, due primarily to the bias modulation in the differential technique. The coupling of $X^-$ to the cavity is made evident by the order of magnitude increase in signal compared to $X^0$ and polarization anisotropy for the $X^-$ signal. The optical response for V polarization is 70 times greater than for H. The $\Delta R$ for $X^0$ shows the anisotropic exchange splitting[23], but the polarization anisotropy in signal strength is essentially gone since $X^0$ is detuned by 6 meV from the cavity resonance. The linewidth of $X^-$ is ~30 μeV, much greater than the 8 μeV linewidth of $X^0$ and ~30 times greater than the expected radiative limit for a QD outside the cavity. We attribute this linewidth increase to the cavity-QD interaction.

In Fig. 2a,b the ΔR for V and H polarizations are plotted for a series of temperatures, showing the behavior as a function of cavity detuning. The $X^-$ feature for V polarization decreases by a factor of 20 at 34 K (1 meV detuning) while the $X^-$ line for H polarization decreases only by a factor of 2, reducing the polarization anisotropy. The $X^-$ linewidth also decreases by as much as 30% with increasing temperature, consistent with a decrease in coupling to the cavity. The linewidth decrease continues until 28K, when the linewidth starts to increase, presumably due to phonon broadening. The ΔR for V polarization is fitted to a model that describes scattering of light from the coupled cavity-QD system, including interference with background reflections (see Supplement). From these fits we obtain a cavity-dot coupling $g_c$ of 25μeV. This relatively small coupling strength is likely due to modest spatial overlap between the QD and cavity mode.

## Spin initialization and measurement

To achieve spin measurement and initialization, a transverse magnetic field is applied that splits the electron and $X^-$ energy levels. This results in the four transitions in Fig. 2c, which can be considered two Λ-type three level systems (the two differing in the $X^-$ spin state). In Fig. 2d, the lower ΔR signal shows the cavity mode with only some small ripples near the expected $X^-$ transitions. This result is an indication of optical pumping[24-26]. When a laser is resonant with one of the transitions, a small change in reflectivity occurs if the system is in the spin state being optically driven, thus measuring the spin state.

But recombination to the other spin state can occur, so the QD system is quickly pumped out of the spin state being driven and into the other, eliminating absorption from the QD. This effect is used to initialize the spin qubit. The four optical transitions are experimentally observed in Figure 2d only when an optical pulse resonant with all four transitions (upper curve) is present to counter the effects of optical pumping. The presence of some residual signal from the QD without the pulse is likely due to incomplete pumping and interference effects.

To better characterize optical initialization, it is useful to moderately detune the QD from the cavity resonance, making the four transitions easier to resolve. Temperature tuning can accomplish this, but optical pumping is defeated at higher temperatures by faster spin relaxation. Instead, we examine a different QD-cavity system (QD-C2) in Fig. 3, in which $X^-$ is 1 meV above the cavity resonance at 7 K. In Fig. 3a $\Delta R$ for V polarization is 3 times larger than for H, indicating the QD is coupled to the cavity. At a magnetic field of 4T (inset of Fig. 3a), the four transitions are observed with strong polarization dependence. We note that in our QDs the polarization axis is randomly oriented, presumably due to valence band mixing[25, 27]. For this particular QD, the axis is close to the cavity orientation, with the outer (inner) transitions predominantly coupling to V (H). For QD-C1 the axis is closer to 45°, so all four transitions couple well to the cavity.

To eliminate the dispersive lineshapes in $\Delta R$ that make it difficult to extract initialization fidelities, we measure differential resonance fluorescence[28, 29] ($\Delta RF$, see supplement) of QD-C2 by exciting with H polarization and detecting V, blocking out the reflected laser. RF provides a nearly background-free alternative to $\Delta R$ as a way to measure the spin state. Instead of measuring a small change in reflectivity when the system is in the spin state being optically driven, photons are only scattered and detected when in that spin state. In Fig. 3b, on the edges of the charge stability region (1.87V and 1.97V) where rapid cotunneling occurs[25], all four transitions appear. Nominally, the outer transitions should not appear for H excitation, but the polarization selection rules are not perfectly aligned with H/V, and the cavity enhances emission for these transitions. As the bias moves toward the center of the charge stability region, the inner transitions disappear due to optical pumping, giving an initialization fidelity of at least 95%. The outer transitions never entirely disappear because they are driven much more weakly with H polarization, and because the pumping rate is significantly slower for these transitions as we will now show.

An important feature of the cavity is that the pumping rate is strongly modified, but in an asymmetric way. This is illustrated in the inset diagrams of Fig. 3c. When driving an outer (V) transition (lower diagram), which is more strongly coupled to the cavity, emission is likely to return the system to its initial spin state, thus giving a slow pumping rate. When driving an inner (H) transition (upper diagram), emission is more likely to change the spin state, thus giving a fast pumping rate. We determine these rates by temporally resolving the bleaching of RF due to optical pumping after a short pulse depolarizes the spin. In Fig. 3c, the optical pumping rates are plotted as functions of laser power for the laser tuned to the inner and outer transitions, with pumping rates for a QD outside the cavity plotted for comparison. The pumping rate at saturation is an order of magnitude higher for the inner transitions than the outer even with moderate detuning. Thus, driving the transitions that are not coupled to the cavity permits fast initialization, and driving the transitions coupled to the cavity is better for spin measurement. For enhanced measurement it is especially advantageous to have well aligned polarization axes and large Purcell enhancement. Optimizing these would give a cycling transition that could permit single shot readout.

## Ultrafast single qubit gates

A functional spin qubit also requires single qubit gates (*i.e.* spin rotations). This is accomplished with short, circularly polarized pulses that couple the two spin states together through one of the two $X^-$ spin states (Fig. 4b)[19, 30-37]. A significant complication is that the cavity only couples to V polarization, and yet the rotation requires circular polarization. We find that spectrally detuning the pulse from the cavity alleviates this issue. The pulse then primarily interacts with the QD directly instead of through the cavity mode.

To demonstrate these coherent rotations, we measure the spin population after two rotation pulses (~13 ps length, ~150μeV bandwidth) delayed with respect to each other by a variable time $\tau$[32]. The resulting Ramsey interference fringes for QD-C1 are displayed in Fig. 4a. The first pulse rotates the Bloch vector near the equator, where it precesses (see inset of Fig. 4a) for time $\tau$ at the Larmor precession frequency. The second pulse rotates the spin up or down depending on the phase, giving rise to oscillations in the spin population. The decay time of the oscillations (~400 ps) is due primarily to dephasing from the fluctuating nuclear polarization[38]. It should be possible to recover the spin coherence using spin echo techniques[39] or by optically suppressing nuclear spin fluctuations[40, 41]. In Fig. 4c, the amplitude of the Ramsey fringes is plotted as a function of rotation pulse power and is indicative of damped Rabi oscillations of the electron spin. The peaks at 3μW and 11μW correspond to rotation pulses with area of π/2 and 3π/2. The troughs at 7μW and 15μW correspond to rotations of π and 2π. The non-zero amplitude of the Ramsey oscillations for two nominal π pulses is due to precession during the pulses that limits the rotation fidelity. Simulations of the Ramsey fringes (displayed in Fig. 4c) give similar behavior due to this effect.

To determine how the cavity affects the pulse rotation, the maximum Ramsey fringe amplitude (*i.e.* for ~$\pi/2$ rotation pulses) is measured as a function of detuning as shown in Fig. 4d. The amplitude is highly asymmetric about the QD transitions at 1301.95meV. For negative pulse detunings, where the pulses are far from the cavity mode, the Ramsey fringes are strong, and for positive detunings, where the pulses are nearly resonant with the cavity mode, the fringes are quite weak. We compare this result to a theoretical model in which the incident laser field is modified by the cavity mode (see Supplement). Figure 4e plots the theoretical fidelity of a single pulse $\pi/2$ spin rotation and the purity of the resulting spin state after the pulse within this model. The calculations show behavior qualitatively similar to the experiment. The poor fidelity near the cavity resonance arises from real excitation of $X^-$ through the cavity, whereupon recombination destroys the purity of the spin state. The increase in fidelity at a detuning of 0.3meV is due to the cavity field driving the system up to $X^-$ and then partially back down to the ground states, preventing recombination. Both measurement and theory show that the fidelity can be quite high with a large negative detuning, such that the pulse is spectrally far away from the cavity.

## Discussion

We have demonstrated the first functional spin qubit coupled to an optical cavity in which optical initialization, control and readout of an electron spin is achieved. Having a long-lived solid state spin qubit coupled to a cavity opens up many areas of research including CQED with a spin degree of freedom, spin-controlled photonics, and quantum networks. This type of system may in fact be used as a node in a quantum network, with optical pulses controlling the emission of photons into nearby waveguides, transmitting quantum states to other nodes[19]. An important step in this direction will be to provide better control over the parameters of the system, including spectral and spatial overlap with the cavity mode, higher Q-factors, and the polarization axis of the QD relative to the cavity mode. Based on the pumping rates measured here, it seems possible with improved Q-factors and coupling strengths to drastically decrease the initialization time to picosecond timescales, significantly improving qubit operation speed. And with the QD polarization axis better aligned to the cavity polarization, the cavity-enhanced transitions should act as very bright cycling transitions that could be used for single-shot readout. Previously the only demonstrated QD spin system that has both fast spin initialization and cycling transitions is the W-system obtained in coupled QDs[26, 29].

We also anticipate the incorporation of multiple spin qubits within a node. This can be achieved using QD molecules in a cavity, where a tunneling interaction provides two qubit gates[35, 36]. Even without tunnel-coupling, multiple QD spin qubits within a cavity can be entangled using pulse sequences that make use of a cavity-induced interaction[42]. These multiple qubits can be used for error correction and form the basis of a quantum repeater or quantum computer.

## Methods

**Sample structure.** The samples are grown by molecular beam epitaxy on n-type GaAs substrates. A 500 nm sacrificial layer of n-$Al_{0.7}Ga_{0.3}As$ is grown on the substrate, followed by 50 nm of n-GaAs, 40 nm of intrinsic GaAs, 3 nm InAs QDs, 50 nm of intrinsic GaAs, and 30 nm of p-GaAs. The QDs are distributed randomly in the growth plane with a density of several QDs per $\mu m^2$. A positive e-beam resist (ZEP520A, Zeon Chemicals Co.) has been used for defining 2D PhCs by Raith 150 e-beam lithography system, followed by $Cl_2$-based ICP etching. A triangular lattice of 70 nm radius holes with a lattice spacing of 242 nm are etched through the epilayer into the AlGaAs, with three missing holes at the center forming an L3 cavity (see Fig. 1a). The AlGaAs underneath each PhC is etched away, leaving a 180 nm thick PhC membrane. Ohmic contact is made to the p-type layer on the surface and to the n-type substrate.

**Measurement techniques.** The sample was mounted on piezo-stages in a magneto-optical cryostat, with the magnetic field oriented parallel to the long dimension of the cavity. A 0.68 NA aspheric lens focused lasers onto the sample and collected PL and reflected laser light. For $\Delta R$ measurements, the bias of the diode was modulated at 1-10 kHz with a square-wave peak-to-peak amplitude of ~200-400 mV, and the modulation of the reflected light was measured with lock-in detection. Small shifts in the cavity position with bias give rise to a $\Delta R$ signal from the cavity that is essentially the derivative with respect to bias, giving rise to dispersive line-shapes.

For Ramsey interference fringe measurements, the cw laser was used to initialize the system and readout the spin state. Pulses from the Ti:Sapphire laser were split into two 13 ps pulses with the opposite circular polarization as the cw laser, so that in detection the pulses were rejected using a polarizer. The cw laser was on in-between the pulses, which may lead to some decoherence. This effect appeared to be negligible compared to effects from nuclear spins.

## Acknowledgements

This work was supported by a Multi-University Research Initiative (US Army Research Office; W911NF0910406) and the US Office of Naval Research. We thank Alex Greilich for valuable contributions in the preliminary stage of this research.

## Author Contributions

All authors were involved in preparing the manuscript. S.G.C., T.M.S., A.S.B., and D.G. conceived and designed the experiments and samples. A.S.B. grew the QD samples. M.K., C.S.K., and A.S.B. processed photonic crystals and gates in the samples. T.M.S., S.G.C., and L.Y. optically characterized the cavities and QDs. S.G.C. performed the differential reflectivity and laser control experiments. D.S., S.E.E., T.L.R., and T.M.S. provided theoretical insight and calculations.## Additional Information

The authors declare no competing financial interests.

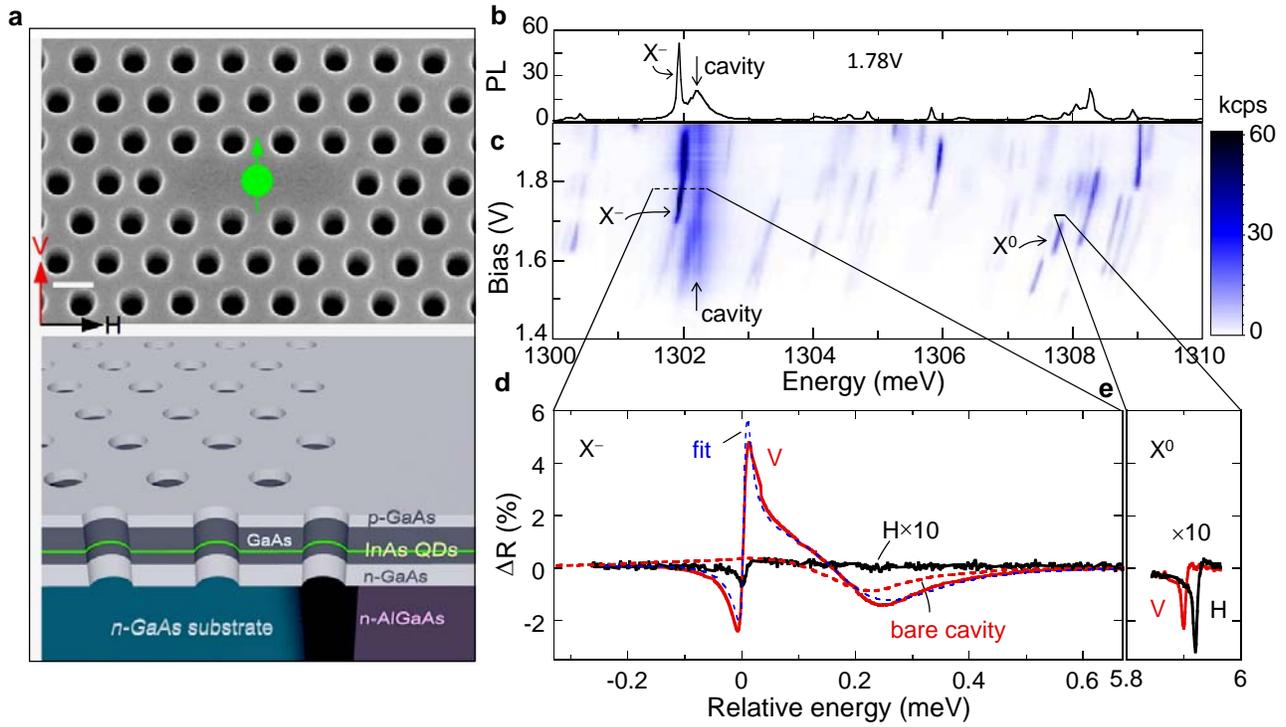

**Figure 1 | Charged QDs in a cavity. a**, (top) Scanning electron micrograph of an L3 PhC cavity with an illustration of an electron spin in the center (the actual QD location is unknown). (bottom) Illustration of the layers that form the p-i-n diode. The scale bar represents 200 nm. **b**, PL of the cavity-QD system at a bias of 1.78V, exciting at 1391 meV. **c**, PL bias map of the cavity-QD system, measured in $10^3$ counts/s (kcps). Horizontal lines correspond to the reflectivity scans below. **d**, Differential reflectivity near $X^-$ at a bias of 1.785V (solid lines), at which the QD is charged with a single electron and at 1.7V (dashed red line), at which the QD is uncharged. H and V correspond to the laser polarization. **e**, Differential reflectivity of $X^0$ at 1.72V. The magnetic field is zero, and the temperature is 7-8 K. The energy scale in d and e is relative to $X^-$ at 1301.93 meV.

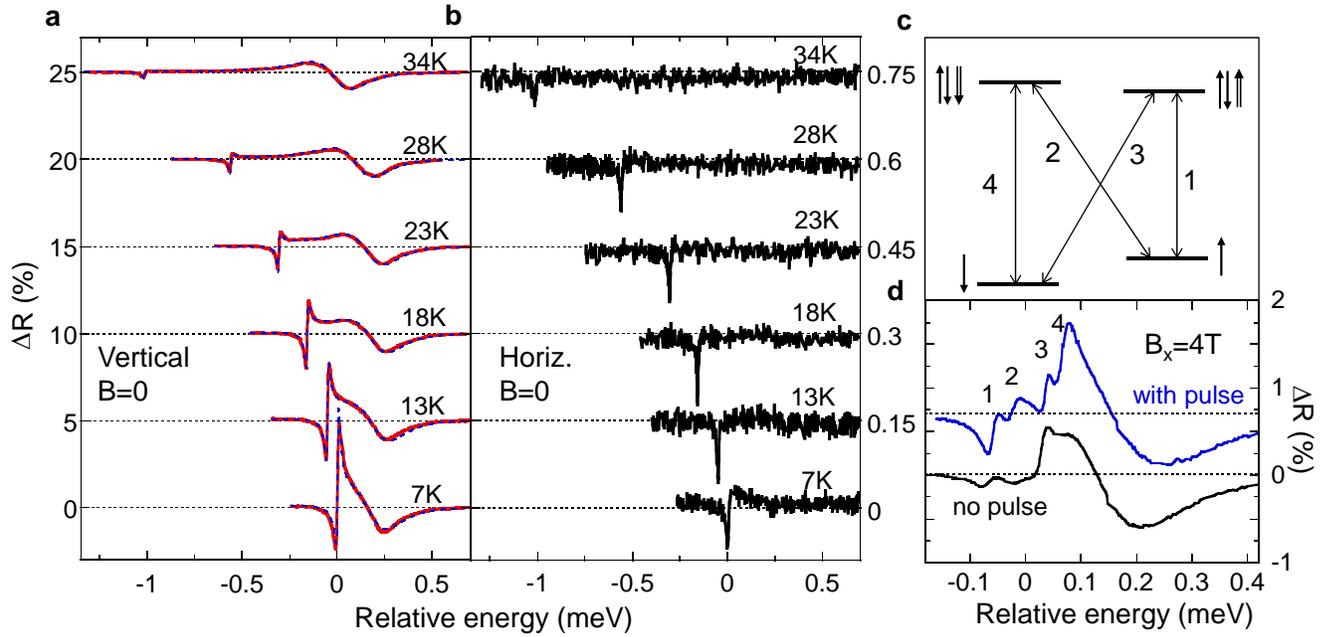

**Figure 2 | Resonant laser spectroscopy. a,b** ΔR for V and H polarizations, respectively, for a series of temperatures, varying the detuning of $X^-$ from the cavity. The dashed blue lines in panel a are fits to the reflectivity. The spectra are vertically offset at each temperature for clarity. **c**, Level diagram showing the electron spin states and $X^-$ spin states in a Voigt magnetic field. Single (double) arrows represent electron (hole) spins. **d**, ΔR at $B_x = 4T$ for circular polarization with (upper curve) and without (lower curve) a short, resonant pulse that defeats optical pumping. The energy scales are relative to $X^-$ at 7 K.

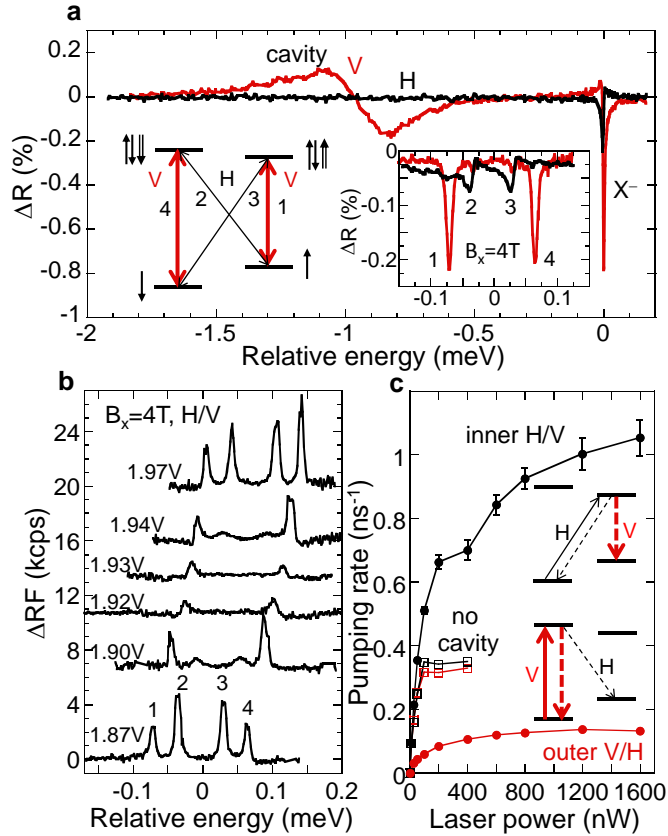

**Figure 3 | Spin initialization/measurement in QD-C2. a,** ΔR for H and V laser polarizations at zero magnetic field and a bias of 1.9V. The inset plot displays ΔR of the X⁻ lines at $B_x$=4 T and 1.87 V. The inset level diagram shows the electron spin states and X⁻ spin states in a Voigt magnetic field. Single (double) arrows represent electron (hole) spins. Thicker red lines indicate cavity enhanced transitions. **b**, ΔRF for H polarized laser, V polarized detection for a series of biases (400 mV modulation) at $B_x$=4 T and laser power of 25 nW. Spectra are vertically offset according to bias. **c**, Spin pumping rates vs. laser power for inner/outer transitions for QD-C2 at 1.93 V and a QD outside the cavity. The laser/detection polarization is H/V (V/H) for inner (outer) transitions. Error bars represent the standard error of the nonlinear least-squares fit to an exponential and are shown when larger than the data points. Inset level diagrams illustrate the effect of cavity enhancement on pumping rates. Solid (dashed) lines indicate the excitation laser (spontaneous emission).

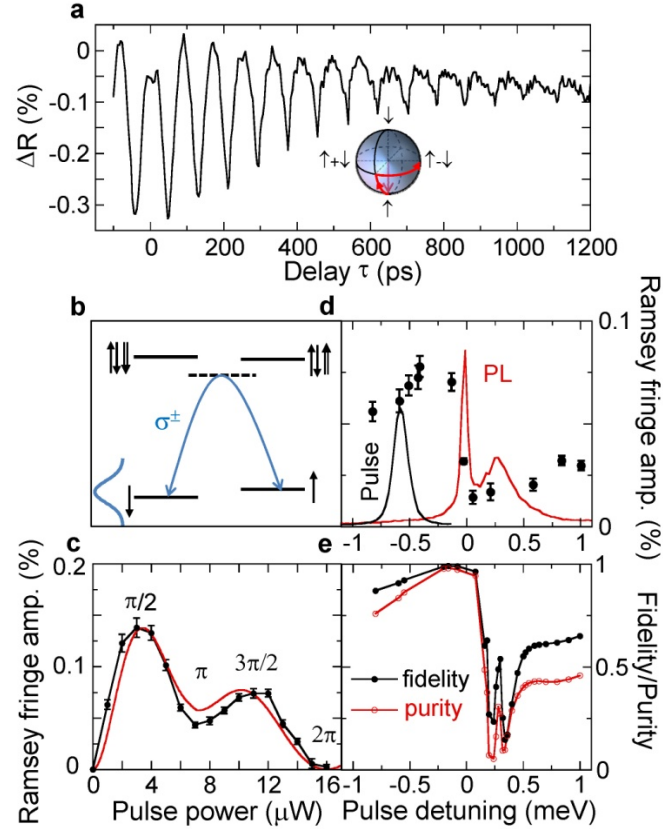

**Figure 4 | Coherent spin rotation. a,** Ramsey interference fringes at $B_x = 2T$ for a pulse detuning of -0.34meV, with the initialization/measurement laser near transition 2. The inset shows a Bloch sphere with rotation of the Bloch vector from the first pulse and precession in-between pulses. **b,** Level diagram showing the coherent Raman process that couples the two electron spin states through an $X^-$ state. The dashed line represents a virtual state. **c,** Fitted amplitude of the Ramsey interference fringes as a function of pulse power, for a detuning of -0.56meV. The red curve is a model calculation. **d,** Maximum Ramsey fringe amplitude (solid circles) as a function of pulse detuning. The PL spectrum (at B=0) and a sample pulse spectrum are displayed for comparison. Error bars in c and d represent the standard error of the nonlinear least-squares fit to a decaying cosine **e,** Theoretical model of the fidelity of the spin rotation and purity of the resulting state as a function of pulse detuning.

# Supplementary Material: Quantum control of a spin qubit coupled to a photonic crystal cavity

Samuel G. Carter[1†], Timothy M. Sweeney[2†], Mijin Kim[3], Chul Soo Kim[1], Dmitry Solenov[2], Sophia E. Economou[1], Thomas L. Reinecke[1], Lily Yang[2], Allan S. Bracker[1], and Daniel Gammon[1]

[1] *Naval Research Laboratory, Washington, DC 20375, USA*
[2] *NRC postdoctoral associate residing at the Naval Research Laboratory, Washington, DC, 20375, USA*
[3] *Sotera Defense Solutions, Inc., Annapolis Junction, MD 20701, USA*
[†] *These authors contributed equally to this work.*

## I. EXPERIMENTAL METHODS

### A. Differential reflectivity

Differential reflectivity ($\Delta R$) is measured by modulating the applied bias with a square wave of peak-to-peak amplitude $V_{mod}$ = 200-400mV and frequency $f_{mod}$ = 1-10 kHz. The biases quoted in the paper are the higher bias, so $\Delta R$ is the reflectivity difference $R(V_{bias}) - R(V_{bias} - V_{mod})$. Typically, at the lower bias the QD has changed charge state or the resonance is Stark-shifted out of range, so $\Delta R$ is only due to the QD at $V_{bias}$. However, the cavity resonance has only a weak dependence on bias, so it essentially shows up as $\frac{dR_{cavity}}{dV_{bias}} V_{mod}$.

Due to the relatively high current through the n-i-p diode (5-20 mA for ~10-20 mm$^2$ devices in the X$^-$ bias range) and the ~50 $\Omega$ output resistance of the function generator applying the modulated bias, the voltage across the sample is different from that generated in the function generator. This has been corrected to give the sample biases listed throughout the main text and supplement. The modulation amplitude $V_{mod}$ is also decreased by roughly a factor of 2 from the nominal 200 or 400 mV.

The dispersive lineshapes of QDs can occur due to the interference between the electric field scattered from the QD and the fields reflected from the surfaces of the PhC membrane[1]. The path differences and phase change upon reflection make the reflectivity sensitive to some combination of the real and imaginary parts of the QD polarization response function. Interference with the cavity reflectivity is also important near the cavity resonance. The reflectivity of the entire cavity-QD system including interface reflection is modeled in section II.A.

## B. Resonance fluorescence (ΔRF)

Resonance fluorescence (RF) simplifies the QD spectra by eliminating background reflections through polarization rejection and only measuring light scattered from the QD. We achieve polarization rejections of $10^5$ - $10^6$ by sending the reflected light through a variable retarder, a linear polarizer and into a single mode fiber, which sends photons to a single photon counting module (SPCM). The variable retarder corrects for any ellipticity induced by cryostat windows or beamsplitters, and the single mode fiber acts to spatially filter the reflected light. This polarization rejection still gives ~$10^3$ counts/s under typical conditions and varies with the laser wavelength. To eliminate this background, we measure the differential signal by modulating the bias at 1 kHz as described for ΔR. In Fig. 1a, RF is shown for the modulated bias, with one channel counting at the high bias, in which X¯ is present, and the other at the low bias, where only background is present. The difference (ΔRF) eliminates the background as displayed in Fig. 1b.

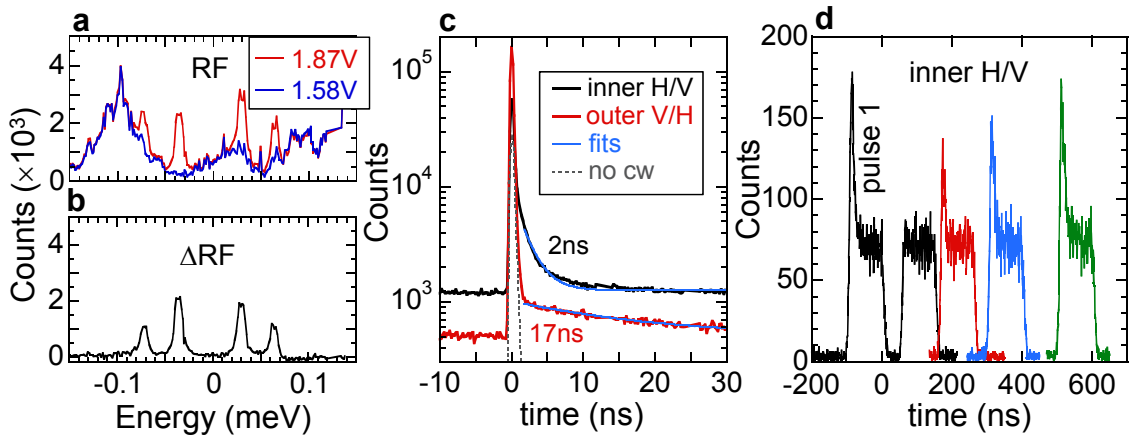

**Figure 1 | Resonance fluorescence on QD-C2. a,** RF spectra at two biases and **b,** the resulting ΔRF spectrum, with $B_x$=4T and a laser power of 25nW. **c,** TCPC of RF at 1.935 V, showing the decay of RF due to optical pumping for the inner and outer transitions. $P_{laser}$= 100 nW and $B_x$=4 T. **d,** TCPC of RF at 1.935 V for two 100 ns pulses, showing the recovery of RF due to spin relaxation. $B_x$=2T, and $P_{laser}$ = 50nW.

The scattered light is not spectrally resolved, so both elastic (Rayleigh) and inelastic (Raman spin flip) scattering is detected. All of this scattered light is considered RF here. In the Voigt magnetic field, the scattered light may be primarily Raman spin flip scattering when the polarization selection rules of the QD are aligned with the excitation/emission polarizations. Rayleigh scattering is suppressed in this case because any single transition will either not couple to the excitation laser or be blocked by the polarizer.

Using time-correlated photon counting (TCPC) with RF, we determine the optical pumping rate and the spin lifetime of QD-C2. To measure optical pumping, the excitation laser is first tuned to an $X^-$ transition at a DC bias (no modulation) in which pumping is observed. A 13 picosecond pulse resonant with all four transitions is used to quickly depolarize the spin state, and RF is then emitted until the pumping is complete. In Fig. 1c, the RF counts as a function of time after the pulse are displayed, showing background before the pulse, a sharp peak from laser scatter and RF from the laser pulse, and exponentially decaying RF from the cw laser. The decay time corresponds to the optical spin pumping rate. A pulse picker increases the pulse repetition period of the 13 ps pulses by a factor of 10 to 123 ns to measure these decay times. Some leakage of pulses every 12.3 ns shows up in the TCPC, so these points are removed for clarity.

The spin relaxation time is measured by observing how long it takes for RF to turn back on after optical pumping has bleached the RF. The cw laser is modulated with an acousto-optic modulator (AOM) into two 100 ns pulses separated by a variable time delay. As displayed in Fig. 1d, the first pulse shows initial RF followed by decay. (This method also measures the pumping rate, but it is limited by the 10 ns turn-on time of the AOM.) For the second pulse, only the background is observed for short pulse delays, indicating no relaxation has occurred. For longer delays, the RF turns back on, indicating relaxation. By measuring the RF recovery as a function of pulse delay, we obtain a spin relaxation time of 265 ns. This value is many orders of magnitude smaller than the millisecond times measured previously for InAs QDs[2]. The short relaxation time appears to be an issue for all the QDs in this n-i-p structure since we have noticed that the bias region where pumping occurs is narrow even for QDs outside of PhC patterns. This may be correlated with the relatively high current in the diode at these biases, which heats up the sample mount by a few degrees.

C. Ramsey fringe measurements

For Ramsey interference fringe measurements, the cw laser was used to initialize the system and readout the spin state. Pulses from the Ti:Sapphire laser were split into two 13 ps pulses with the opposite circular polarization as the cw laser, so that in detection the pulses were rejected using a polarizer. The cw laser is on all the time, continuously measuring the spin state and simultaneously reinitializing the system to the spin state opposite of that being driven[3, 4]. When the sequence of pulses changes the spin state, a change in the ΔR signal occurs until the system is reinitialized. We assume that the system is close to fully initialized before the next pulse sequence ~12 ns later, a good assumption given the pumping rate measurements presented. Having the cw laser on during the pulse sequence may lead to some decoherence. However, this effect appears to be negligible compared to effects from nuclear spins since the fringe decay rate is not sensitive to the cw laser power in this range.

The Ramsey fringe measurements as a function of detuning, displayed in Fig. 4d of the main text, were actually performed for a series of pulse powers. For pulses significantly detuned from $X^-$, the pulse power needed to obtain a $\pi/2$ spin rotation pulse increases with detuning, so Ramsey fringe traces were taken at several powers in order to obtain the maximum Ramsey fringe amplitude. When the pulses are overlapping with $X^-$, the pulses should be designed to drive the system up to the trion and back down to the ground states (a $2\pi$ optical pulse) to avoid leaving population in the trion state[5]. This gives a spin rotation angle of $\pi$ at zero detuning, which decreases with detuning. For simplicity, we take the maximum Ramsey fringe signal as a function of pulse power in this regime as well.

## II. THEORETICAL MODELS
### A. Reflectivity

In order to understand the reflectivity data presented in Fig. 1 and 2 of the main text, we follow the experimental procedure and calculate the $\Delta R$ signal. The light reflected from the sample contains only a small contribution coming from the cavity mode (CM) and the $X^-$ transition in the quantum dot (QD). Most of the reflected signal is due to scattering off "inactive" parts of the sample, such as the microcavity membrane surfaces, substrate, etc. In order to extract the reflectivity signal originating from the QD and the CM we calculate the differential reflectivity signal, $\Delta R$, a difference between reflectivity with, $R$, and without, $R_0$, the active dot. In what follows we derive $\Delta R$ taking into account scattering of light from the QD, the CM, and the background. Coupling between the cavity and QD is also taken into account.

The reflectivity, as measured by the detector, is proportional to the square of the electric field $\mathbf{E}$ at the detector. This amplitude, up to a constant factor, can be obtained from the equation

$$E_i(\mathbf{r},\omega) = E_i^0(\mathbf{r},\omega) + G_L(\mathbf{r},\omega) \sum_{j=x,y,z} \int d\omega' A_R^{ij}(\omega,\omega') E_j^0(0,\omega'), \qquad (1)$$

where $\mathbf{E}^0$ is the electric field amplitude of the incoming light, before scattering, $G_L(\mathbf{r},\omega) \sim 1/r$ is light propagator, and $A_R^{ij}(\omega,\omega')$ is the scattering kernel (or amplitude) that describes reflection of $\mathbf{E}^0$ from the sample. The scattering amplitude is determined by a set of microscopic processes constituting the reflection from the sample. It separates into three parts

$$A_R = A_D + A_C + A_{BG}. \qquad (2)$$

The first part in Eq. (2), $A_D$, represents scattering from QD. This scattering involves the following microscopic chain of events: the incoming light is absorbed by the QD (trion transition X⁻) with amplitude $g$; the trion persists for some time [with amplitude $G_D^0(t-t')$]; and the trion recombines, emitting a photon, with amplitude $g$. We obtain the overall amplitude

$$A_D(\omega) \sim g G_D(\omega) g, \qquad G_D(\omega) \to G_D^0(\omega) = [\omega - \omega_D + i\Gamma_D]^{-1}, \qquad (3)$$

where $\omega_D$ is the X⁻ frequency.

The second part of Eq. (2) involves the scattering from the CM. It is due to the overlap of electromagnetic field inside and outside of the cavity, such that vacuum photon can tunnel into (and out of) the CM with probability $\gamma$. In this case we obtain

$$A_C(\omega) \sim \gamma G_C(\omega) \gamma, \qquad G_C(\omega) \to G_C^0(\omega) = [\omega - \omega_C + i\Gamma_C]^{-1}, \qquad (4)$$

where $\omega_C$ is the frequency of the CM. In both cases, higher-order (in $g$ and $\gamma$) terms due to multiple subsequent transitions to and from QD and CM can be lumped into respective decay rates $\Gamma_D$ and $\Gamma_C$. The coupling $g_C$ between the QD and CM allows for multiple QD-CM absorption-emission events that can be summed into the respective dressed Green's function,

$$G_C(\omega) = \cfrac{1}{\omega - \omega_C + i\Gamma_C - \cfrac{|g_C|^2}{\omega - \omega_D + i\Gamma_D}}, \qquad (5)$$

and a similar expression for $G_D$.

Finally, the last part of Eq. (2) is the amplitude of background scattering from surfaces. This scattering, $A_{BG} = A e^{i\varphi}$, is essentially frequency independent compared to the scattering from the QD and the CM.

When the polarization of incoming and collected light is perpendicular, or horizontal (H), to that of the CM, light only couples to the QD directly. As the result $R_H \sim A_D + A_{BG}$. When the polarization of the incoming and collected light matches that of the CM, *i.e.* is vertical ($V$), the reflectivity signal comes from all three sources mentioned above. The $\Delta R$ signal is stronger in the latter case indicating that $\gamma$ is much larger than $g$. As a result, the contribution of $A_D$ to V polarization signal can be neglected and the overall amplitude becomes

$$A_{R,V}(\mathbf{r},\omega) \sim e^{i\varphi} + G_C(\omega)|\gamma|^2 / A \qquad (6)$$

In order to eliminate large background signal we follow the experiment and subtract the reflectivity signal without the charged QD. Note, however, that the frequency of the cavity in this case is shifted slightly, by $\delta\omega$. We obtain

$$\Delta R_V \sim \left| e^{i\varphi} + G_C(\omega)|\gamma|^2 / A \right|^2 - \left| e^{i\varphi} + G_C^0(\omega+\delta\omega)|\gamma|^2 / A \right|^2 \qquad (7)$$

Finally, we expand $\Delta R_V$ in powers of $1/A$ retaining only the leading 0th-order term. We also expand in $\delta\omega$ up to the first order, *i.e.* $\Delta R_V = \Delta R_V^0 + \delta\omega \Delta R_V^1 + O(\delta\omega^2)$.

We fit Eq. (7) to the vertical polarization $\Delta R$ data by first identifying $\omega_C^0$ and $\omega_D^0$ ($\omega_{C/D}^0 \neq \omega_{C/D}$) at which $\Delta R_V = 0$ for each temperature. The remaining fitting parameters $g_C$, $\Gamma_C$, $\Gamma_D$, $\varphi$, and the overall amplitude are the same for all the data. At 34K $\omega_C - \omega_D \gg g_C$ and the $\Delta R$ due to the cavity can be fitted separately, producing the same results for $\Gamma_C$ and $\varphi$ (as well as the amplitude). The values obtained from the fit are $\Gamma_C = 172\,\mu eV$, $\Gamma_D = 5.2\,\mu eV$, $|g_C| = 24.9\,\mu eV$, $\phi = 1.13\,\text{rad}$, $\omega_D = (1301.94,$ 1301.88, 1301.77, 1301.62, 1301.37, 1300.92) meV, and $\omega_C = (1302.13, 1302.14, 1302.13, 1302.11,$ 1302.05, 1301.93) meV.

As we see from Fig. 2 of the main text, the above analysis provides an accurate estimate of the QD-CM coupling constant $g_C$. At the same time, it does not directly provide the Purcell spontaneous emission enhancement factor. The latter, however, can be estimated based on the values of $g_C$ and $\Gamma_C$. The Purcell factor is a ratio $\Gamma'/\Gamma_0$, where $\Gamma' = g_C^2/\Gamma_C$ is the spontaneous emission rate into the cavity mode (in resonance), and $\Gamma_0$ is the spontaneous emission rate into vacuum (about $0.5\,\mu eV$). From this we obtain an estimate $\Gamma'/\Gamma_0 \sim 7$. Note that $\Gamma_D \neq \Gamma_0$ since the $X^-$ transition linewidth is not at the radiative limit. An indirect indication of the Purcell effect can also be gained by comparing the 30 µeV linewidth of $X^-$ near the CM to the 8 µeV linewidth of $X^0$, as mentioned in the main text. This broadening is due to the QD-CM interaction and also appears in the model.

### B. Ramsey fringes

The data in Fig. 3c of the main text is modeled by numerically integrating the optical Bloch equations for the three-level system in response to a sequence of two pulses (hyperbolic secant, 11.5 ps FWHM) detuned 0.56 meV below the QD. Decay of the coherence at high pulse powers is modeled by multiplying the calculated fringe amplitude by a decaying exponential. When the pulse length is not significantly shorter than the Larmor precession period, precession during the pulse limits rotation fidelity. This leads to the non-zero signal for two nominal π-pulses. When the pulse length is roughly 1/3 of the precession period, two nominal π-pulses actually give rise to a maximum in the Ramsey fringe amplitude instead of an expected minimum. This led to an error in the assignment of rotation angles in Fig. 4e of Ref. 4, in which the angles should be twice as large as labeled.

### C. Spin rotation with a cavity.

To describe the spin rotation process by an off-resonant pulse in the presence of the cavity, we use a semiclassical formalism, where the QD spin is treated quantum mechanically and the cavity is treated classically. This is a reasonable approximation because we are in a moderate coupling and Q regime. In this regime, the QD sees a pulse that is modified by the cavity. The total electric field can be written as $\mathbf{E}_{tot}(t) = \mathbf{E}_{laser}(t) + \hat{e}\int d\omega\, G(\omega)\tilde{E}_{laser}(\omega)\,e^{-i\omega t}$, where the tilde denotes the Fourier transform of the laser field, $\hat{e}$ is a vector along the polarization of the cavity, and $G(\omega)$ is the symmetrized Green's function describing the photonic mode in the cavity[6], with $G(\omega)/\omega_C = G_C^{(0)}(\omega) + G_C^{(0)}(-\omega)$. The cavity mode frequency is $\omega_C$. We use a hyperbolic secant for the laser temporal shape, with the pulse length set at each detuning to give a π/2 rotation in the absence of the cavity[5]. Precession during the pulse is ignored since errors of this sort can be corrected. There is an overall adjustable parameter multiplying the Green's function, which we adjust to mimic the behavior in the experiment (the depth and width of the fidelity/purity plots as functions of detuning).

In the equations of motion for the density matrix of the QD spin, besides the external pulse the optical spontaneous emission rate enters as well. We modify that rate compared to its value in free space $\Gamma_0$ to take into account that the QD can emit into the cavity. We use a heuristic expression which takes into account the detuning between cavity and QD: $\Gamma_0 \to \Gamma_0 + g_c^2 \Gamma_c / (\Gamma_c^2 + \delta^2)$, where $\delta$ is the detuning between the cavity and the QD. A similar formula has been used previously (see, for example, Ref. 7).